# The effect of turbulence on drifting snow sublimation

Zhengshi Wang[1, 2, 3, 4], Ning Huang[1, 2,*], Thomas Pähtz[5, 6,†]

[1] Key Laboratory of Mechanics on Environment and Disaster in Western China, Ministry of Education, Lanzhou University, Lanzhou 730000, People's Republic of China

[2] Department of Mechanics, School of Civil Engineering and Mechanics, Lanzhou University, Lanzhou 730000, People's Republic of China

[3] State Key Laboratory of Aerodynamics, China Aerodynamics Research and Development Center, Mianyang, Sichuan 621000, China

[4] Computational Aerodynamics Institute, China Aerodynamics Research and Development Center, Mianyang, Sichuan 621000, China

[5] Institute of Port, Coastal and Offshore Engineering, Ocean College, Zhejiang University, 866 Yu Hang Tang Road, 310058 Hangzhou, China

[6] State Key Laboratory of Satellite Ocean Environment Dynamics, Second Institute of Oceanography, 36 North Baochu Road, 310012 Hangzhou, China

**Abstract**

[1] Sublimation of drifting snow, which is significant for the balances of mass and energy of the polar ice sheet, is a complex physical process with intercoupling between ice crystals, wind field, temperature, and moisture. Here a three-dimensional drifting snow sublimation model in a turbulent boundary layer is proposed. In contrast to most previous models, it takes into account turbulent diffusion of moisture from lower to higher elevations, allowing the air humidity near the surface to be undersaturated and thus sublimation to occur. From simulations with this model, we find that snow sublimation in the saltation layer near the surface dominates overall snow sublimation, despite an only marginal departure from humidity saturation (< 1%), because of a large particle concentration.

[†]Corresponding author: tpaehtz@gmail.com

[*]Corresponding author: huangn@lzu.edu.cn

# 1 Introduction

[2] As one of the basic elements of Earth's environment, snow is widely distributed in high and cold areas, and its temporal and spatial evolution is crucial for the hydrological cycle [*Zhou et al.*, 2014]. The mass and energy balance of ice shelves, which plays a key role in climate change [*Cess and Yagai*, 1991; *Solomon et al.*, 2007], is largely affected by 'drifting snow', describing the transport of snow by wind, a phenomenon common in high and cold regions [*Huang et al.*, 2016; *Schmidt*, 1972]. In fact, drifting snow may slow down global warming due to heat-trapping during snow crystal sublimation [*Solomon et al.*, 2007], which occurs when the humidity of air is unsaturated [*Schmidt*, 1972]. As the sublimation rate of drifting snow particles is generally larger than that of ground snow due to the larger area of exposure [*Dai and Huang*, 2014; *Huang et al.*, 2016], and since the accumulated time of drifting snow accounts for up to 1/3 of the duration of winter in the polar region [*Mahesh et al.*, 2003; *Mann et al.*, 2000], understanding the drifting snow phenomenon can play an important role in modeling mass and energy variations of the ice sheet.

[3] However, although numerous experimental [*Bintanja*, 2001b; *Mann et al.*, 2000; *Schmidt*, 1982; *Wever et al.*, 2009] and theoretical studies [*Déry and Yau*, 2002; *Groot Zwaaftink et al.*, 2011; *Vionnet et al.*, 2013; *Xiao et al.*, 2000] on drifting snow sublimation have been conducted, sublimation of drifting snow within the saltation layer (the near-surface layer in which most particles move in characteristic ballistic hops) is only rarely investigated [*Dai and Huang*, 2014; *Huang et al.*, 2016; *Sharma et al.*, 2018] as the air humidity near the surface saturates rapidly when drifting snow occurs because of a high snow concentration [*Bintanja*, 2001a; b; *Mann et al.*, 2000]. However, because of moisture transportation, there are always slight deviations from the saturated state, which give rise to snow sublimation, and the impact of such

deviations has not yet been quantified.

[4] Here we explore drifting snow sublimation in a natural turbulent atmospheric boundary layer and the impact of moisture transport by turbulent motions using the large eddy simulation (LES) model of the Advanced Regional Prediction System (ARPS) [*Xue et al.*, 2001], where the drifting snow in the turbulent boundary layer is simulated using a model that accounts for the main particle transport processes [*Lämmel et al.*, 2017]. The sublimation of mid-air snow particles is modeled by the sublimation rate formula of ice crystals [*Thorpe and Mason*, 1966], and its feedback effects on the air temperature and moisture is considered through introducing temperature and moisture source terms into the corresponding diffusion equations.

**2 Generation of drifting snow**

[5] The fluid governing equations of the LES model are described in the supplementary material. In the turbulent boundary layer, drifting snow occurs when the local shear stress $\tau$ is larger than the critical value $\tau_t$. The rate of aerodynamic entrainment can be expressed as [*Anderson and Haff*, 1991]:

$$n_a = \eta(\tau - \tau_t) \tag{1}$$

where $\eta = 1.5 / 8\pi d_p^2$ [*Doorschot and Lehning*, 2002], and $\tau_t = 0.2^2 g \langle d_p \rangle (\rho_p - \rho)$ [*Clifton et al.*, 2006], in which $g$ is the gravitational acceleration, and $\rho_p$ and $d_p$ are, respectively, the density and diameter of snow particle. Note that the mass density of saltating snow particles is very close to that of ice cubes because most branches of snowflakes are knocked out in their impacts with the surface [*Comola et al.*, 2019; *Groot Zwaaftink et al.*, 2011]. The density of air $\rho = p(1 - q/(\varepsilon + q))(1 + q)/(R_d T)$ obeys the state equation of moist air, $R_d$ is the gas constant of dry air, and $T$ and $q$ are, respectively, the temperature and the water vapor mixing ratio of air. Furthermore,

$\varepsilon \equiv R_d / R_v$, where $R_v$ is the gas constant of water vapor. Note that aerodynamic entrainment is dominated by splash entrainment in the steady state [*Kok et al.*, 2012], thus, the exact values of the parameters in equation (1) do not have a significant effect on the final results.

[6] Particle trajectories are calculated using the Lagrangian particle tracking method, and the wind speed at the particle location is determined by the linear interpolation of surrounding grid points. Treating the snow grain as a sphere and subjecting it only to the gravity and drag force, the motion equation of a snow particle can be written as [*Anderson and Haff*, 1988]:

$$\frac{du_{pi}}{dt} = \frac{3\nu V_{ri}}{4\rho_p (d_p)^2} C_D Re_p + g_i (1 - \frac{\rho}{\rho_p}) \qquad (2)$$

where $u_p$ is the particle velocity, $t$ the time, $\nu$ the kinematic viscosity of air, $V_r$ the relative speed between snow particle and local wind, and $Re_p = d_p V_r / \nu$ is the particle Reynolds number. Note that we use the center point of snow particle to calculate the grid location. The drag coefficient $C_D$ is calculated as [*Bagnold*, 1941]:

$$C_D = \frac{24}{Re_p} + \frac{6}{1 + Re_p^2} + 0.4 \qquad (3)$$

[7] To describe grain-bed interactions, we use the grain scale-dependent splash function of *Lämmel et al.* [2017] because it has been derived mostly from first physical principles and because, in contrast to other available splash functions [*Comola and Lehning*, 2017; *Kok and Renno*, 2009], it takes into account the energy partition occurring in subsequent binary collisions between bed particles [*Crassous et al.*, 2007; *Ho et al.*, 2012]. The rebound probability of an impacting particle is related to its impact velocity $v_{in}$ and incident angle $\theta_{in}$, which can be expressed as [*Lämmel et al.*, 2017]:

$$P_{reb}(v_{in}) = 1 - \frac{1 + \ln \xi}{\xi} \qquad (4)$$

with

$$\xi \equiv \frac{2\sqrt{2}(\alpha+\beta)^2 \hat{d}_p \theta_{in}}{\beta^2 \left(\theta_{in} + \sqrt{2g\hat{d}_{in}}/|v_{in}|\right)^2} \qquad (5)$$

where $\hat{d}_{in}$ and $\hat{d}_p$ are the diameters of the impacting and bed particles, respectively, which are both normalized by $(d_{in} + d_p)/2$. Furthermore, $\alpha = (1+\varepsilon_n)/(1+\mu) - 1$ and $\beta = 1 - (2/7)(1+\varepsilon_t)/(1+\mu)$ are parameters related to the microscopic normal and tangential restitution coefficient $\varepsilon_n$ and $\varepsilon_t$, respectively, of the material and $\mu = \varepsilon_n d_p^3 / (d_p^3 + \varepsilon_n \langle d_p \rangle^3)$ is the modified mass ratio of the collision partners.

[8] According to *Higa et al.* [1998], the normal restitution coefficient of ice typically has a constant value $e_{qe}$ in the quasi-elastic regime and a tendency to decrease in the inelastic regime, which can be described by a piecewise function:

$$\varepsilon_n = \begin{cases} e_{qe} & v_n < v_c \\ e_{qe}\left(\dfrac{v_n}{v_c}\right)^{-\log(v_n/v_c)} & v_n \geq v_c \end{cases} \qquad (6)$$

in which $e_{qe} = \exp\left(-\pi\eta/\sqrt{1-\eta^2}\right)$, with $\eta = 0.027(1+d^3)^{0.1}(1+d)^{0.2}(d_{in}/0.05)^{-0.5}$, $d \equiv d_{in}/d_p$, $v_n$ is the normal relative velocity of the collision partners, and $v_c$ is the critical impact velocity for the transition from the quasi-elastic region to the inelastic regime, which can be expressed as:

$$v_c = v_0 (1+d^3)^{3/4} (1+d)^{-7/4} \exp\left(\frac{c}{2RT}\right) \left(\frac{d_{in}}{d_0}\right)^{-1/2} \qquad T \geq 229K \qquad (7)$$

where $v_0$ and $d_0$ are constant parameters (all parameter values are specified in Table

1).

[9] The total restitution coefficient of a rebounding particle does not vary significantly, and thus is set to be its mean value [*Lämmel et al.*, 2017]:

$$\bar{e} = \beta - (\beta^2 - \alpha^2) \langle d_p \rangle \theta_{in} / (2\beta) \tag{8}$$

A rebound angle $\theta_{re}$ is picked from the following probability distribution function [*Lämmel et al.*, 2017]:

$$P(\theta_{re}|\theta_{in}) = \begin{cases} \dfrac{\beta^2 (\theta_{in} + \theta_{re})}{(\alpha+\beta)^2 \sqrt{3} \langle d_p \rangle \theta_{in}} \ln \dfrac{2(\alpha+\beta)^2 \sqrt{3} \langle d_p \rangle \theta_{in}}{\beta^2 (\theta_{in} + \theta_{re})^2}, & 0 < \dfrac{\beta(\theta_{in} + \theta_{re})}{(\alpha+\beta)\sqrt{3}\langle d_p \rangle \theta_{in}} < \sqrt{2}, \\ 0, & else, \end{cases} \tag{9}$$

[10] The impacting particle also generates a number of low energy ejection particles $N_e$ [*Lämmel et al.*, 2017]:

$$N_e = \gamma \frac{(1-\bar{e}^2) E_{in}}{2(\bar{E}_e + \phi_c)} erfc\left(\frac{\ln E_m - \mu_e}{\sqrt{2}\sigma_e}\right) \tag{10}$$

where $\gamma = 0.06$ is a dimensionless constant, $E_{in} = m_{in} v_{in}^2 / 2$ is the kinetic energy of impact particle and $E_m = \langle m_e \rangle g \langle d_p \rangle$ is the minimum transferred energy for a bed particle to be counted as ejecta, $m_{in}$ and $m_e$ are the mass of impact and ejection particle, respectively, and $\bar{E}_e = E_m \left[(1-\bar{e}^2) E_{in}/E_m\right]^{1-(2-\ln 2)\ln 2}$ is the average energy of ejection particles. In a manner analogous to *Comola and Lehning* [2017], equation (10) has been modified from *Lämmel et al.* [2017] to take into account the mean cohesion energy $\phi_c$ between bed particles. Furthermore, $\mu_e$ and $\sigma_e$ are mean value and variance of the energy of ejected particles, respectively, which can be expressed as:

$$\mu_e = \ln\left[(1-\bar{e}^2) E_{in}\right] - \lambda \ln 2 \tag{11}$$

$$\sigma_e = \sqrt{\lambda} \ln 2 \tag{12}$$

with

$$\lambda = 2\ln\left[\left(1-\overline{e}^2\right)E_{in}/E_m\right] \tag{13}$$

[11] Then, the energy of ejection particles can be obtained from the log-normal distribution as following [*Lämmel et al.*, 2017]:

$$P\left(E_e | E_{in}\right) = \frac{1}{\sqrt{2\pi}\sigma_e E_e}\exp\left[-\frac{\left(\ln E_e - \mu_e\right)^2}{2\sigma_e^2}\right], \tag{14}$$

and the ejection angle follows the exponential distribution with a mean value of 50° [*Kok and Renno*, 2009].

[12] Because the turbulence frequency spectrum of atmospheric boundary layers and thus the diffusion features are different from that of wind tunnels, we simulate two configurations. The simulation domain of configuration 1 (wind tunnel) has the dimensions $2\times 1\times 1$ m, and the uniform horizontal grid space is 0.02 m. The simulation domain of configuration 2 (field) has the dimensions $10\times 5\times 16$ m, and the uniform horizontal grid space is 0.1 m. Near the bed surface, where most snow particles are transported, the mesh is denser in the vertical direction, and the smallest grid sizes are 0.002 m and 0.01 m for configuration 1 and 2, respectively. For both configurations, the lateral boundary conditions for wind field, humidity, temperature, and snow particles are periodic. The top boundary is set as the stress-free boundary condition, and the bottom is a rigid wall (at which the moisture and heat fluxes are zero). The values of the other model parameters are the same for both configurations and summarized in Table 1.

[13] The wind field is initialized with the standard logarithmic wind profile:

$$u(z) = \frac{u_*}{\kappa}\ln\left(\frac{z}{z_0}\right) \tag{15}$$

where $u_*$ is the friction velocity and $\kappa = 0.41$ is the Karman constant, and $z_0$ is the surface roughness (configuration 1: $z_0 = 3.0e-5$ m; configuration 2: $z_0 = 1.0e-4$ m). This defines the pressure gradient $u_*^2/H$ (where $H$ is the thickness of the boundary layer) and the wall friction $f_w = -\tau_w A_{cell}$ (where $A_{cell}$ is the area of the grid and $\tau_w = \rho u_*^2$ is the local wall shear stress), both of which are kept constant with time. The development time for the turbulent boundary layer is set equal to six times of the large-eddy turn over time $T_*$ ($T_* \equiv H/u_*$), and drifting snow begins after the turbulent boundary layer is fully developed. In the case that a particle reaches the top, it rebounds without energy loss. However, we do not observe particles that exceed a hop height of about 6.0 m.

[14] The snow particle size distribution follows the gamma function [*Nishimura and Nemoto*, 2005; *Schmidt*, 1982]:

$$P(d_p) = \frac{1}{\beta_p^{\alpha_p} \Gamma(\alpha_p)} d_p^{\alpha_p - 1} \exp(-\frac{d_p}{\beta_p}) \tag{16}$$

where $\alpha_p$ and $\beta_p$ are the shape and scale parameter, respectively. During the calculation, the diameter of each takeoff particle is picked from above function. The particle size distribution for configuration 1 resembles that of the wind tunnel experiments by *Sugiura et al.* [1998] ($\alpha_p = 6$ and $\beta_p = 50$, with $d_{10} \approx 150\ \mu m$, $d_{50} \approx 300\ \mu m$ and $d_{90} \approx 480\ \mu m$, respectively), while the particle size distribution for configuration 2 resembles that of the field experiments by *Schmidt* [1982] ($\alpha_p = 5$ and $\beta_p = 50$, with $d_{10} \approx 125\ \mu m$, $d_{50} \approx 250\ \mu m$ and $d_{90} \approx 400\ \mu m$, respectively).

[15] The initial air temperature is uniformly 258.15 K. At the top boundary, the relative humidity is set constant (configuration 1: 60%; configuration 2: 70%). The humidity profile becomes steady once the sublimation rate of drifting snow particles matches the moisture flux at the upper boundary, as governed by the upper relative

humidity boundary condition.

**3 Results and discussion**

[16] One feature of the LES model is that turbulence eddies larger than the grid scale can be resolved, which provides a means to explore the diffusion features of moisture and heat in the turbulent flow, and which is more sophisticated than modeling diffusion through the diffusion equation [*Huang et al.*, 2016; *Wever et al.*, 2009]. Drifting snow in the turbulent boundary layer is generated through entrainment of surface snow directly by the wind and/or the bombardment of snow particles that are already transported by the wind ('splash') [*Clifton et al.*, 2006; *Sugiura and Maeno*, 2000]. Hence, when turning on the boundary layer flow, the number of snow particles increases until an equilibrium is reached, in which snow entrainment is balanced by snow deposition at the surface.

[17] For both configurations, the simulated steady state does not depend on the choice of the initial humidity profile. This steady state is reached when the production of water vapor balances turbulent diffusion. All simulation results are averaged over a period of 120 s during the steady state. Steady state profiles of transport rate and relative humidity obtained from our simulations for both configurations are shown in Figure 1(a-c) and compared with wind tunnel and field experiments from the literature [*Bintanja*, 2001b; *Sugiura et al.*, 1998; *Wever et al.*, 2009]. Both the associated equilibrium snow transport flux profile and humidity profile are consistent with observations. Furthermore, Figure 1(d) compares the overall simulated sublimation rate for various friction velocities $u_*$ with the measurements by *Bintanja* [2001b]. The data in Figure 1 (a) and (b) were measured in a wind tunnel with snow samples that were stored in a cold room at -15 $°C$ for an extended period, and those in Figure 1 (c) and (d) are both field observations at the Swedish station Svea under natural

drifting snow. Further model validations are shown in the supplementary material. The observed agreement between simulations and measurements gives us confidence that the numerical model captures the essential physics of drifting snow sublimation. However, there is no certainty about the appropriateness of the model assumptions given that the experimental data used for the model evaluation cover only the region above the saltation layer.

[18] Numerous measurements and model predictions have shown that a nearly saturated layer exists near the surface [*Bintanja*, 2001a; b; *Déry and Yau*, 2002; *Mann et al.*, 2000], thus, sublimation within this layer is believed to be negligible. To test this hypothesis, we identify the locations $z_x$ at which the relative humidity exceeds certain threshold values x (e.g., for all $z < z_{99\%}$, the relative humidity is larger than 99%). Figure 2(a) shows that the thickness of two representative layers ($z_{99\%}$ and $z_{95\%}$) well exceeds the mean saltation height $z_s$ (defined as the elevation below which 50% of the mass flux takes place [*Kok et al.*, 2012]). However, despite an only marginal departure from humidity saturation, the vast majority of overall snow sublimation takes place within these layers, while sublimation underneath the mean saltation layer height accounts for more than half of the overall sublimation. The reason for this counter-intuitive behavior is that the particle concentration decays exponentially with height [*Pähtz and Durán*, 2018], which means that even a very slight departure from humidity saturation near the surface can have a strong effect compared with highly unsaturated layers on top where much less snow particles are transported. Note that we carried out test simulations without moisture transportation, in which case the humidity near the surface is fully saturated and snow sublimation thus does not take place there. Also, modest changes of the values of the model parameters, such as the mass density of snow, do not have an effect on the quality of

the results.

[19] Because turbulent diffusion does not only drive moisture transport but also tends to keep transported snow particles suspended in air, we now estimate the contribution of suspended snow particle transport to overall transport in our drifting snow simulations (configuration 2). The standard way to do this is by examining the transport height of snow particles, with particles transported in lower elevations being classified as saltating and particles in larger elevations being classified as suspended [*Bagnold*, 1941]. However, also particles in lower elevations can be supported by turbulent diffusion, which is why we decided for a different estimation of particle suspension: the ratio between the hop time $t_s$ and the hop time $(v_{z0} - v_{zf})/g$ as it would be in the absence of turbulent fluctuations of the vertical wind velocity under pure Stokes drag, where $v_{z0}$ and $v_{zf}$ are the vertical particle velocity at the beginning and end of the hop. In fact, for ratios $t_s/\left[(v_{z0} - v_{zf})/g\right]$ close to unity, particles are far from being suspended, whereas for large ratios, suspension plays a significant role. As shown in Figure 3, there is, indeed, a large peak at unity in the distribution of this ratio. Nonetheless, about 20% of all particles have hop times ratios that are 24.5 and larger, which means that suspended particle transport is considerable. However, note that the limited vertical dimension of the simulation domain does not allow us to simulate deep suspension clouds that sometimes occur in nature (e.g., in Antarctica [*Mahesh et al.*, 2003; *Mann et al.*, 2000]).

[20] Recently *Sharma et al.* [2018] showed that the Thorpe-Mason model, which we use to model snow sublimation, underestimates the sublimation rate when the residence time of snow particles is too short assuming that air and ice have the same initial temperature. This underestimation affects large particles more strongly because

of their small residence time, which are those particles that tend to be transported in lower layers. Hence, the contribution of near-surface layers to overall sublimation is expected to be even stronger than what we find from our model.

## 4 Conclusion

[21] In this study, drifting snow sublimation in the turbulent boundary layer has been investigated using a numerical model that considers the coupling effect between snow particles, wind field, and the effects of ice phase transition on air temperature and moisture. In particular, we carried out simulations to study the impact of turbulent diffusion on snow sublimation occurring near the surface where the relative humidity is very high (e.g., > 99%). We find that snow sublimation near the surface is the main contributor to overall snow sublimation and much larger than that of higher elevations where the humidity is highly unsaturated. Our results imply that snow sublimation near the surface, neglected in most previous studies, likely plays an important role for the mass and energy balances of snow cover in high and cold regions.

**Acknowledgements**

[22] This work is supported by the Second Tibetan Plateau Scientific Expedition and Research Program (2019QZKK0206), the State Key Program of the National Natural Science Foundation of China (41931179), the National Natural Science Foundation of China (11772143, 41371034, 11750410687), the National Key Research and Development Program of China (2016YFC0500900), and the CARDC Fundamental and Frontier Technology Research Fund (FFTRF-2017-08, FFTRF-2017-09). Open source code of Advanced Regional Prediction System (ARPS) is available at http://www.caps.ou.edu/ARPS/. Initial field data and original data of figures are available at https://doi.org/10.6084/m9.figshare.6270038.v1.

**Figures:**

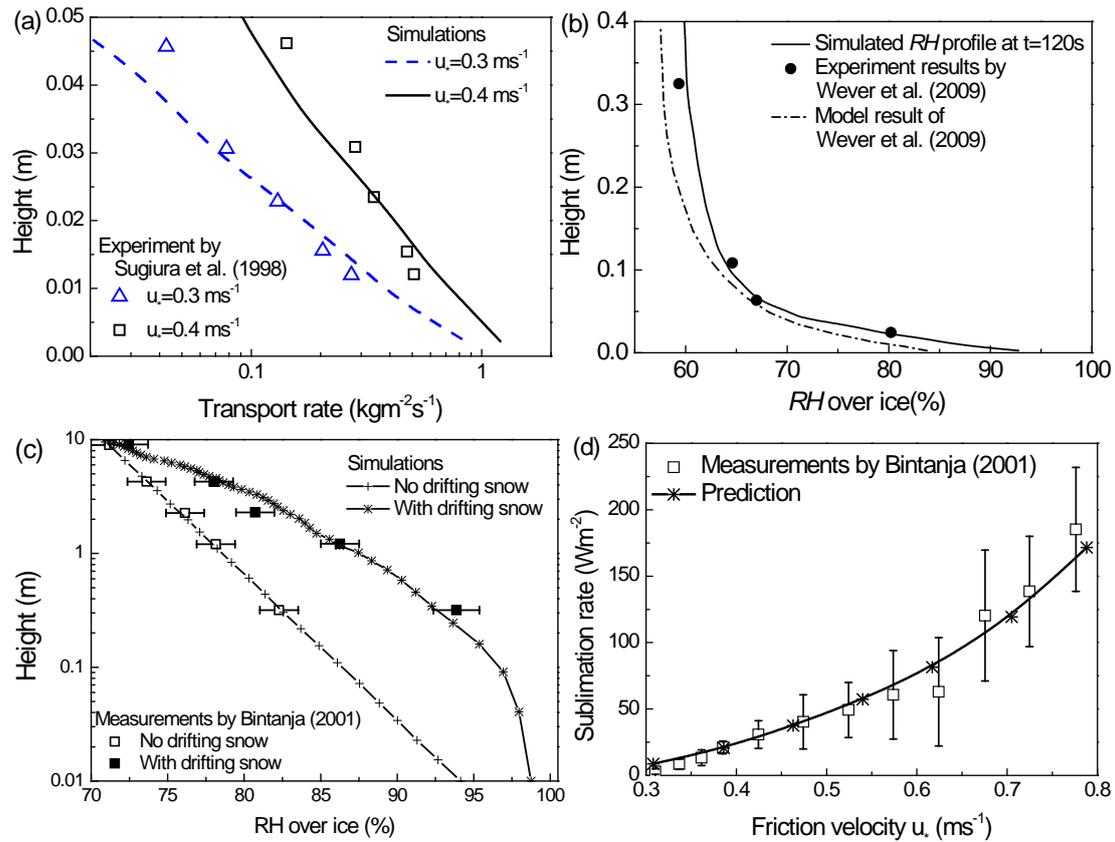

Figure 1. Comparisons of simulated results and observations: (a) Snow transport flux profile and (b) height profile of the relative air humidity at steady state of drifting snow for configuration 1, and (c) height profile of the relative air humidity and (d) variation of sublimation rate versus friction velocity for configuration 2.

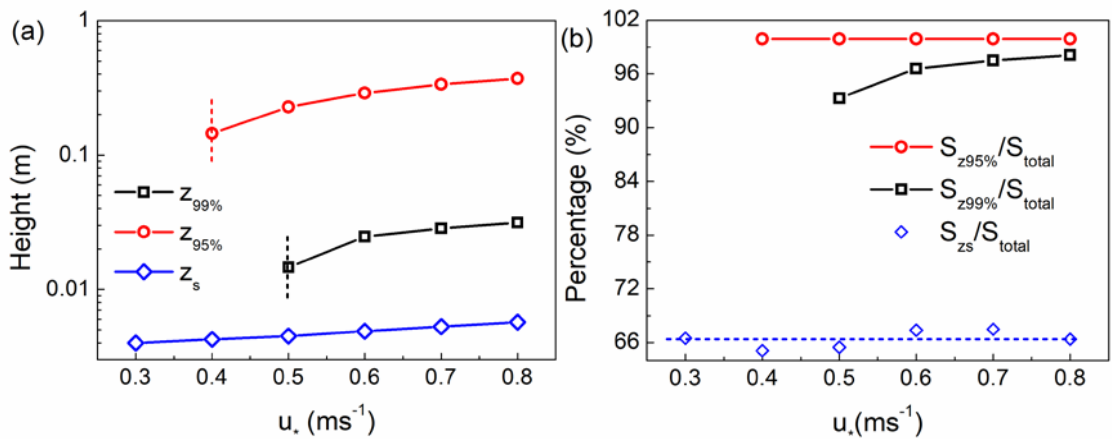

Figure 2. (a) Evolution of $z_{99\%}$, $z_{95\%}$ and the saltation height $z_s$ with friction

velocity and (b) ratio between sublimation occurring for $z < z_{99\%}$, $z_{95\%}$, $z_s$ and total sublimation for configuration 2.

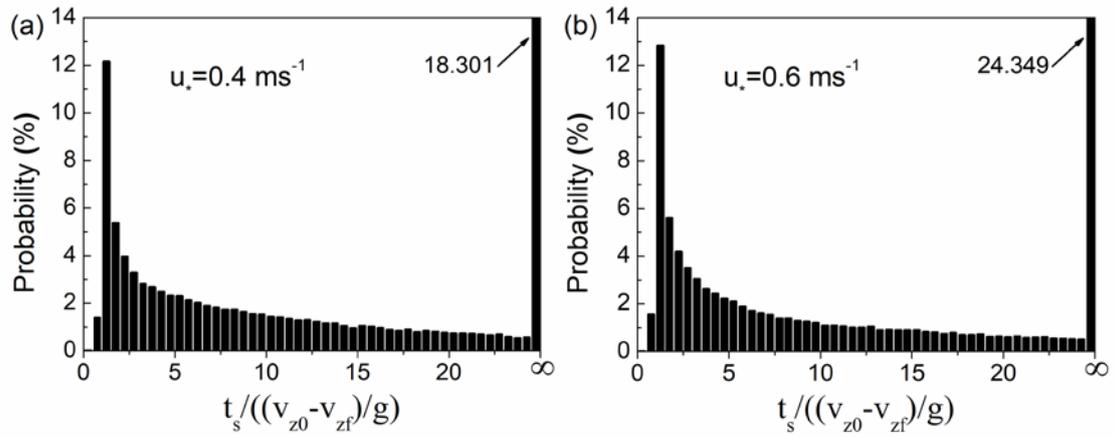

Figure 3. Probability distribution of the ratio of particle hop time $t_s$ to its expected value $(v_{z0} - v_{zf})/g$ in the absence of suspension for configuration 2 for a friction of (a) 0.4 ms$^{-1}$ and (b) 0.6 ms$^{-1}$. The right-most vertical bars in (a) and (b) show the total fraction of particles with a hop time ratio between 24.5 and infinity.

**Tables:**

| Parameter | Description | Value (unit) | Reference |
|---|---|---|---|
| $R_d$ | Gas constant of dry air | 287 ($Jkg^{-1}K^{-1}$) | - |
| $R_v$ | Gas constant of water vapor | 461 ($Jkg^{-1}K^{-1}$) | - |
| $\gamma$ | Damping coefficient | 0.5 | - |
| $g$ | Gravitational acceleration | 9.8 ($ms^{-2}$) | - |
| $Pr$ | Prandtl number | 0.75 | - |
| $\rho_p$ | Density of snow particle | 912 ($kgm^{-3}$) | - |
| $\nu$ | Kinematic viscosity of air | 1.5e-5 ($m^2s^{-1}$) | - |
| $L_s$ | Latent heat of sublimation of ice | 2.84e6 ($m^2s^{-2}$) | - |
| $C_p$ | Specific heat of air | 1004 ($m^2s^{-2}K^{-1}$) | - |
| $K$ | Thermal conductivity of air | 0.0227 ($Wm^{-1}K^{-1}$) | - |
| $a, b, c, f_a, f_b$ | Parameters of saturated vapor pressure respect to ice surface | $a$=611.15, $b$=22.452, $c$=0.6, $f_a$=1.0003, $f_b$=4.18e-8 | - |
| $\phi_c$ | Cohesion energy applied on a particle on the bed | 1.0e-10 (J) | [*Gauer*, 2001; *Groot Zwaaftink et al.*, 2011] |
| $v_0, d_0$ | Parameters of microscopic normal restitution coefficients | $v_0 = 5.72e-4$, $d_0 = 0.03$ | [*Higa et al.*, 1998] |
| $\varepsilon_t$ | Microscopic tangential restitution coefficients | -0.85 | [*Supulver et al.*, 1995] |

Table 1. Parameters and constants used in this work.